\documentclass[twocolumn,showpacs,amsmath,nofootinbib,pra,aps,amssymb,longbibliography,superscriptaddress]{revtex4-2} 
\usepackage[T1]{fontenc}
\usepackage[utf8]{inputenc}
\usepackage{times}
\usepackage{color} 
\usepackage{array}
\usepackage{amssymb,amsmath}
\usepackage{amsbsy}
\usepackage[pdftex]{graphicx} 
\usepackage{bm}
\usepackage{float}
\usepackage{dcolumn}

\usepackage[unicode,breaklinks]{hyperref}
\hypersetup{
    unicode=true,
    plainpages=false, 
    colorlinks=true,
    linkcolor=blue,
    citecolor=blue,
    filecolor=black,
    urlcolor=blue
}
\usepackage{url}
\usepackage{verbatim}

\newcommand{\add}{a_{dd}}
\newcommand{\edd}{\epsilon_{dd}}
\newcommand{\br}{\mathbf{r}}
\newcommand{\bq}{\mathbf{q}}
\newcommand{\bu}{\mathbf{u}}
\newcommand{\bx}{\mathbf{x}}

\newcommand\gammaQF{\gamma_\mathrm{QF}}

\begin{document}
 
\title{Excitations of a two-dimensional supersolid} 
\author{E.~Poli}
 \affiliation{
     Universit\"{a}t Innsbruck, Fakult\"{a}t f\"{u}r Mathematik, Informatik und Physik, Institut f\"{u}r Experimentalphysik, 6020 Innsbruck, Austria
 }
	\author{D.~Baillie} 
\affiliation{%
	Dodd-Walls Centre for Photonic and Quantum Technologies, Dunedin 9054, New Zealand}
\affiliation{Department of Physics, University of Otago, Dunedin 9016, New Zealand}
\author{F.~Ferlaino} 
 \affiliation{
     Universit\"{a}t Innsbruck, Fakult\"{a}t f\"{u}r Mathematik, Informatik und Physik, Institut f\"{u}r Experimentalphysik, 6020 Innsbruck, Austria
 }
\affiliation{Institut f\"ur Quantenoptik und Quanteninformation, \"Osterreichische Akademie der Wissenschaften, Technikerstra{\ss}e 21a, 6020 Innsbruck, Austria}	 
	\author{P.~B.~Blakie} 
\affiliation{%
	Dodd-Walls Centre for Photonic and Quantum Technologies, Dunedin 9054, New Zealand}
\affiliation{Department of Physics, University of Otago, Dunedin 9016, New Zealand}
\date{\today} 
\begin{abstract}  
We present a theoretical study of the excitations of the two-dimensional supersolid state of a Bose-Einstein condensate with either dipole-dipole interactions or soft-core interactions.  This supersolid state has three gapless excitation branches arising from the spontaneously broken continuous symmetries. Two of these branches are related to longitudinal sound waves, similar to those in one-dimensional supersolids.  The third branch is a transverse wave arising from the non-zero shear modulus of the two-dimensional crystal. We present the results of numerical calculations for the excitations and dynamic structure factor characterising the density fluctuations, and study their behavior across the discontinuous superfluid to supersolid transition.
We show that the speeds of sound are described by a hydrodynamic theory that incorporates generalized elastic parameters, including the shear modulus. Furthermore, we establish that dipolar and soft-core supersolids manifest distinct characteristics, falling into the bulk incompressible and rigid lattice limits, respectively.
\end{abstract} 

\maketitle
\section{Introduction}
A supersolid is a state of matter in which phase and translational symmetries are broken.  
Here we refer to a $D$-dimensional supersolid as being a system that spontaneously develops a $D$-dimensional crystal structure whilst still exhibiting superfluidity.  For this system it is expected that the broken symmetries will lead to $(D+1)$-Nambu-Goldstone modes that manifest as gapless excitation branches \cite{Watanabe2012a}.  The experimental production of supersolid states in atomic gases \cite{Leonard2017a,Li2017a,Tanzi2019a,Bottcher2019a,Chomaz2019a} has generated interest in their properties, including their  excitation spectra. The  $D=1$ case \cite{Natale2019a,Roccuzzo2019a,Tanzi2019b,Guo2019a,Petter2021a,Hofmann2021a,Ilg2023a} has two gapless excitation branches of longitudinal character, which are referred to as first and second sound\footnote{These associated excitation branches are also commonly referred to as density and phase branches. The second sound is associated to a reduced superfluidity (i.e.~normal component) arising from the spontaneously broken translational symmetry (see \cite{Yoo2010a,Hofmann2021a,Sindik2024a}).}.  A  $D=2$ supersolid has recently been been produced using a dipolar Bose gas in an oblate shaped trapping potential \cite{Norcia2021a,Bland2022a}. This system has an rich phase diagram with different ground state crystal patterns separated by first order transitions \cite{Zhang2019a,Poli2021a,Hertkorn2021b,Zhang2021a,Ripley2023a}, and is an interesting system for considering the interplay of crystalline order with vortices \cite{Gallemi2020a,Ancilotto2021a,Sindik2022a,Poli2023a}.

Here we study the excitation spectrum of a  $D=2$ supersolid. We consider the case of a zero-temperature gas in the thermodynamic limit, which admits a well-defined band structure allowing us to investigate the gapless modes in detail. We show that a hydrodynamic theory provides a precise description of the excitations in terms of a set of underlying elastic parameters. Supersolids with $D>1$ exhibit a shear modulus, characterising the stiffness of the crystal to  transverse (i.e.~shear) deformations. This also manifests as a new gapless branch of transverse excitations, where the motion of the crystal is perpendicular to the direction of wave propagation. This is in contrast to the other two gapless branches which are longitudinal.

To illustrate the properties of $D=2$ supersolids we present results for two systems. These  systems differ in the relative importance of elastic and compressibility effects, and exhibit different behavior for first and second sound across the transition. The first system is a dipolar Bose-Einstein condensate (BEC) in a planar trap, where the atoms are confined along the direction that dipoles are polarized and free in the perpendicular plane. The second system is a 2D BEC of atoms interacting with a finite-ranged soft-core interaction.  
The excitations of both systems can be obtained by numerically solving the Bogoliubov-de Gennes (BdG) equations. While the excitation spectrum for the 2D soft-core system has been previously studied \cite{Saccani2012a,Macri2013a,Kunimi2012a}, we present the first results for 2D dipolar excitations in the thermodynamic regime.  
The thermodynamic limit has the advantage over finite (harmonically trapped) system  studies (e.g.~see \cite{Hertkorn2021a,Hertkorn2021b}), because the in-plane quasimomentum is a good quantum number, leading to well-defined excitation bands that allow a clear interpretation of their properties (e.g.~speeds of sound).  For both systems we develop a hydrodynamic model, involving five elastic parameters that  we determine from ground state calculations.  
We show that the hydrodynamic model provides an accurate description of the speeds of sound determined by numerical calculation of the BdG excitations, and provides insight into the origin of the different behavior of the two system across the transition, and deep into the crystal regime. While both systems studied  are not directly comparable (e.g.,~the dipolar BEC is a three-dimensional system, while the soft-core BEC is 2D) by studying both we reveal the general applicability of  hydrodynamic theory to 2D supersolids.

The outline of the paper is as follows. In Sec.~\ref{Sec:systems} and Appendices \ref{Sec:ModelDBEC} and \ref{Sec:ModelSCBEC} we describe the two systems we study in this work, and their transition to a supersolid state with a triangular crystalline structure. We present results for the excitations and dynamic structure factors determined by numerical solutions of the BdG equations in Sec.~\ref{Sec:ExcitationResults}. In Sec.~\ref{Sec:Hydro} we outline the hydrodynamic theory for the $D=2$ supersolid. We discuss how the relevant elastic parameters that appear in this theory can be determined from ground state calculations. The hydrodynamic predictions for the speeds of sound are compared to those obtained from the BdG calculations. We then discuss the key parameters distinguishing the behavior of the dipolar and soft-core systems and identify the two relevant limits of the hydrodynamic results to describe these systems. Finally, we conclude in Sec.~\ref{Sec:Conclusions}.

\section{Supersolid systems}\label{Sec:systems}
Here we introduce the two systems   examined in this work. Both systems are described by Hamiltonians that are translationally invariant in the $xy$-plane. The ground state phase diagram depends on the average atomic areal density, $\rho$, and  various microscopic parameters (e.g.~interactions), with  2D crystalline ground states occurring in appropriate parameter regimes.  

\begin{figure}[htbp]
	\centering
	\includegraphics[width=3.4in]{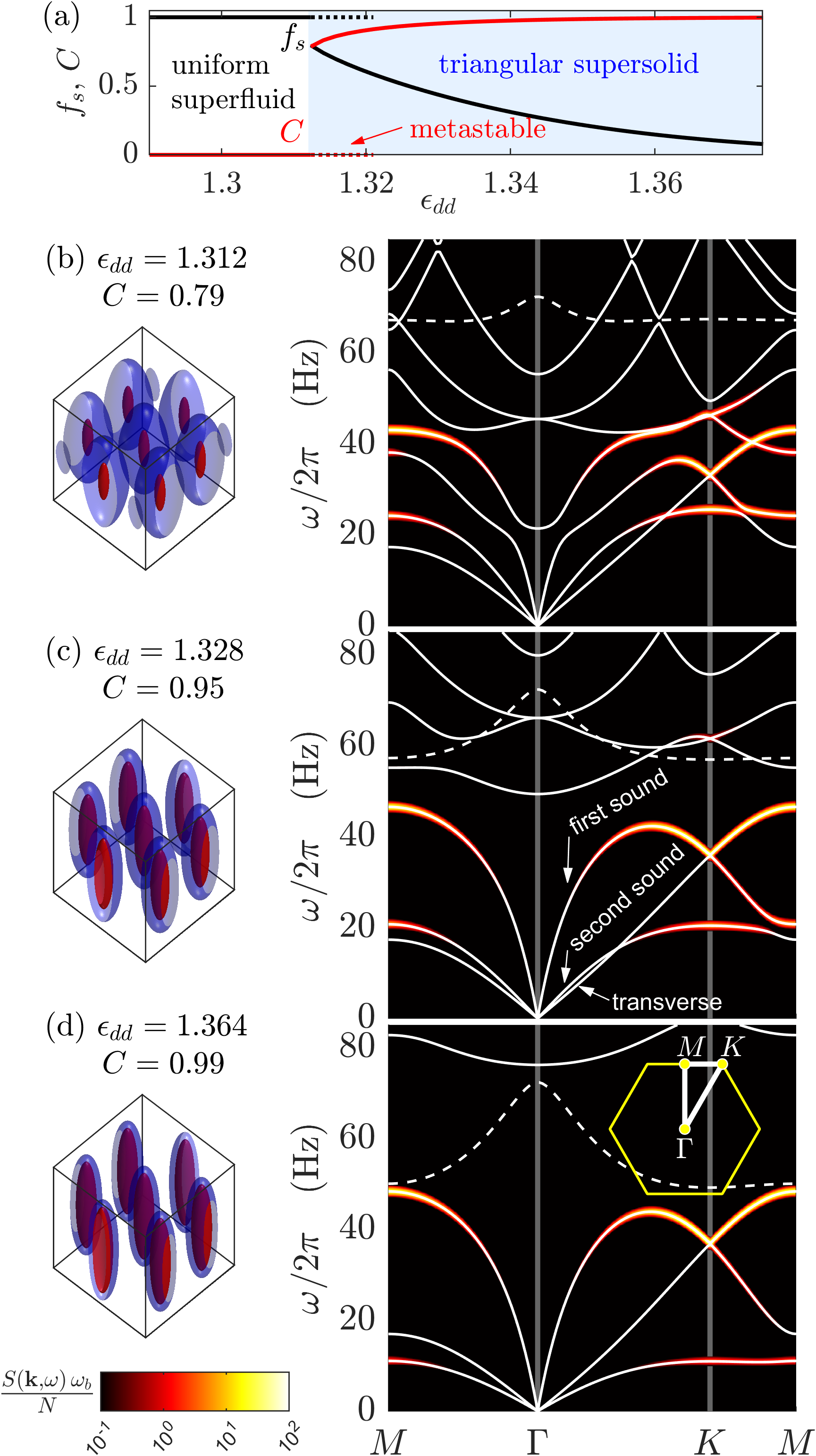}
	\caption{(a) The superfluid fraction $f_s$ and density contrast across the first-order uniform superfluid to supersolid phase transition.
	(b)-(d) Ground state density (left) and excitation spectra along the three symmetry directions of the Brillouin zone (right) [see inset to subplot (d)] for  a planar dipolar gas at three values of $\epsilon_{dd}$. Density isosurfaces at  $0.55\times10^{20}$m$^{-3}$ (blue) and $3\times10^{20}$m$^{-3}$ (red) and shown in a box of size $11\,\mu$m$\times11\,\mu$m$\times14\,\mu$m.  The heat map image  shows the dynamical structure factor frequency broadened by $\omega_b=10^{-2}\omega_z$. 
    The excitation spectra are shown as white lines, dashed lines being odd axial modes which don't contribute to the structure factor.
	Results for $^{164}$Dy atoms with an average areal density of $\rho=0.04/a_{dd}^2$  and axial confinement of $\omega_z/2\pi=72.4\,$Hz. The superfluid fraction is discussed in Sec.~\ref{Sec:ElasticParams}.
	\label{bandstructure}}
\end{figure}

The first system we introduce is a dipolar BEC of highly magnetic atoms under axial harmonic confinement with angular frequency $\omega_z$, but with the atoms free to move in the $xy$-plane. The atoms interact by a short ranged contact interaction with $s$-wave scattering length $a_s$ and the long-ranged dipole-dipole interaction, characterized by the dipole length $a_{dd}$. In this system quantum fluctuation effects become important in the dipole dominant regime, $\epsilon_{dd}\equiv a_{dd}/a_s>1$, with the fluctuations able to  stabilise the condensate from mechanical collapse \cite{Ferrier-Barbut2016a,Wachtler2016a,Bisset2016a}. This system is well-described by an extended meanfield theory with details given in Appendix \ref{Sec:ModelDBEC}. Here we consider a BEC of $^{164}$Dy atoms with $\omega_z/2\pi=72.4\,$Hz, and $a_{dd}=130.8\,a_0$, to be comparable to the phase diagrams produced for this system in Refs.~\cite{Zhang2019a,Ripley2023a}. The system is constrained to have mean areal density $\rho$, which serves as a thermodynamic parameter. We choose to present results for $\rho=0.04/a_{dd}^2\approx8.3\times10^{14}$m$^{-2}$, which is well below the critical point density of $\rho_c\approx0.098/a_{dd}^2$, and is comparable to the densities used in experiments \cite{Norcia2021a}. Under these conditions the nature of the ground state depends on the value of $a_s$, which is controlled in experiments using Feshbach resonances.
For our choice of density the ground state is uniform for $\epsilon_{dd}<1.31$ (i.e.~$a_s>99.7a_0$) and is a triangular crystal for $\epsilon_{dd}\ge1.31$.  
We show examples of these states in Fig.~\ref{bandstructure}. As $\epsilon_{dd}$ increases the overlap between unit cells decreases and at each lattice site a filament shaped droplet (elongated in the dipole direction $z$) forms. It is useful to characterise the strength of the modulation of the density in the $xy$-plane by the density contrast defined as
\begin{align}
C=\frac{\varrho_{\max}-\varrho_{\min}}{\varrho_{\max}+\varrho_{\min}},
\end{align}
where $\varrho_{\max}$ and $\varrho_{\min}$ are the maximum and minimum of the areal density $\varrho(x,y)$ of the ground state.  The density contrast as a function of $\epsilon_{dd}$ is shown in  Fig.~\ref{bandstructure}(a), revealing the discontinuous character of the transition. In these results we have extended the uniform superfluid state beyond the transition point where it is a metastable state.  However, at $\epsilon_{dd}\approx 1.32$ ($a_s\approx99.0a_0$), a roton excitation softens and the uniform state is dynamically unstable.

\begin{figure}[htbp]
	\centering
	\includegraphics[width=3.4in]{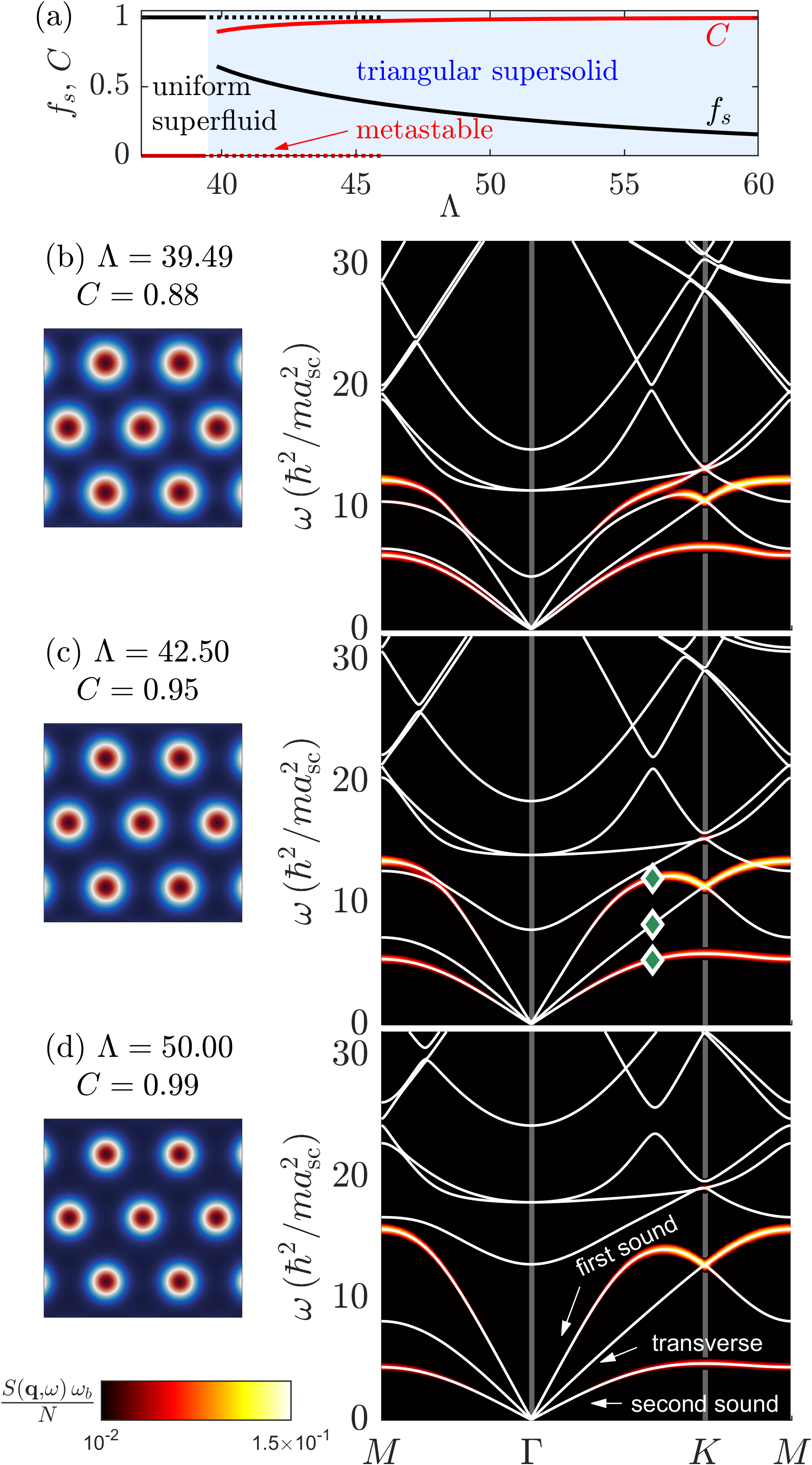}
	\caption{(a) The superfluid fraction and density contrast across the first-order uniform superfluid to supersolid phase transition.
	(b)-(d) Ground state density (left) and excitation spectra along the three symmetry directions of the Brillouin zone (right) for a 2D soft-core system at three values of $\Lambda$. The ground state densities are shown in a box of size $4\,a_\mathrm{sc}\times4\,a_\mathrm{sc}$. The excitation spectra are shown as white lines, on top of a heat map image showing the dynamical structure factor frequency broadened by $\omega_b=0.35\omega_0$. The three symmetry directions of the Brillouin zone are the same as the ones depicted in the inset of Fig.~\ref{bandstructure}(d).  {The three green diamonds mark the  position of the modes plotted in Fig.\,\ref{dynamics_softcore} in an equivalent symmetry direction.}
		\label{bandstructure_softcore}}
\end{figure}

The second system  is a 2D BEC of atoms that interact via the soft-core potential of strength $U_0$ and range $a_{\mathrm{sc}}$.  
This system has been extensively studied as supersolid model (e.g.~see \cite{Pomeau1994a,Josserand2007a,Sepulveda2010a,Saccani2012a,Kunimi2012a,Macri2013a,Cinti2014a,Prestipino2018a}). Schemes have been proposed to realize soft-core interactions in ultra-cold atom experiments \cite{Henkel2010a,Cinti2010a}, but there has been limited reported experimental activity in the regime relevant to supersolidity thus far.  
We consider the system in a regime well-described by meanfield theory, with further details given in Appendix \ref{Sec:ModelSCBEC}. 
The phase diagram for the 2D soft-core model depends on the single dimensionless parameter $\Lambda= {m\pi \rho a_{\mathrm{sc}}^4U_0}/{\hbar^2}$. The melting value $\Lambda_m= 39.49$ separates uniform superfluid states (for $\Lambda<\Lambda_m$) from triangular crystal states.  We show examples of these states in Fig.~\ref{bandstructure_softcore}, and the behavior of the density contrast as a function of $\Lambda$ is shown in Fig.~\ref{bandstructure_softcore}(a). Similar to the dipolar system, the transition from unmodulated to modulates states also occurs discontinuously as $\Lambda$ changes, with the unmodulated states remaining metastable until the roton completely softens at $\Lambda = 46.30$.  
 
For both systems, we find the ground state by minimizing the energy density, $\mathcal{E}$,  i.e.~the energy per unit area. It is convenient to write this  $\mathcal{E}(\rho;\mathbf{v},\mathbf{a}_1,\mathbf{a}_2)$, being a function of $\rho$, the superfluid velocity\footnote{We are interested in stationary superfluids, but introduce  $\mathbf{v}$ to define the superfluid fraction.} $\mathbf{v}$    and $\{\mathbf{a}_1,\mathbf{a}_2\}$, being the direct lattice vectors of the crystal. The ground state for average density $\rho$ is obtained by minimising the energy density for $v=0$ with respect to the lattice constant\footnote{The ground state configuration for both models is a triangular lattice, thus we can take  $\mathbf{a}_1=a \hat{\mathbf{x}}$, and $\mathbf{a}_2=  \frac{1}{2}a \hat{\mathbf{x}}+\frac{\sqrt{3}}{2}a \hat{\mathbf{y}}$, with lattice constant $a$.}. Additional details are given in Appendices \ref{Sec:ModelDBEC} and \ref{Sec:ModelSCBEC}.

\section{Excitation results}\label{Sec:ExcitationResults}
 The excitation spectrum can be determined by linearizing around the ground state $\psi(\mathbf{x})$ with a time-dependent ansatz of the form
 \begin{align}
 \Psi(\mathbf{x},t)=&e^{-i\mu t/\hbar}\left[\psi(\mathbf{x})+\sum_{\nu \mathbf{q}}\left\{c_{\nu\mathbf{q}}u_{\nu\mathbf{q}}(\mathbf{x})e^{-i\omega_{\nu\mathbf{q}}t} \right.\right. \nonumber \\ 
 &\left.\left. -
 c_{\nu\mathbf{q}}^*v^*_{\nu\mathbf{q}}(\mathbf{x})e^{i\omega^*_{\nu\mathbf{q}}t}
 \right\}\right],\label{psipert}
 \end{align}
 where $\mu$ is the ground state chemical potential, $c_{\nu\mathbf{q}}$ are the expansion amplitudes,  and $\nu$  and $\mathbf{q}$ are the band index and planar quasimomentum of the excitation, respectively. Here $\{u_{\nu\mathbf{q}}(\mathbf{x}),v_{\nu\mathbf{q}}(\mathbf{x})\}$ are the quasiparticle amplitudes, with respective energies $\hbar\omega_{\nu\mathbf{q}}$, and these are determined by solving the BdG equations (see Appendix \ref{Sec:BdG}, and Refs.\cite{Baillie2017a,Macri2013a}).

In Figs.~\ref{bandstructure} and \ref{bandstructure_softcore} we show results for the excitation spectra for the dipolar and soft-core models, respectively. These results are shown along the symmetry directions of the first Brillouin zone. For both sets of results the case shown in subplot (b) is close to the phase transition, whereas subplots  (c) and (d) are for states with higher values of the density contrast.  
The excitations are shown as solid white lines on top of the dynamic structure factor, $S(\mathbf{q},\omega)$, which is obtained as
\begin{align}
S(\mathbf{q},\omega)=\sum_\nu \left|\int \!d\mathbf{x}\,(u^*_{\nu\mathbf{q}}-v^*_{\nu\mathbf{q}})e^{i\mathbf{q}\cdot\mathbf{x}}\psi\right|^2\delta(\omega-\omega_{\nu\mathbf{q}}).
\end{align}
The dynamic structure factor reveals the density fluctuations associated with the excitations, notably $S(\mathbf{q},\omega)$ vanishes for excitations that do not affect the density. 

\begin{figure}[htbp]
	\centering
	\includegraphics[width=3.4in]{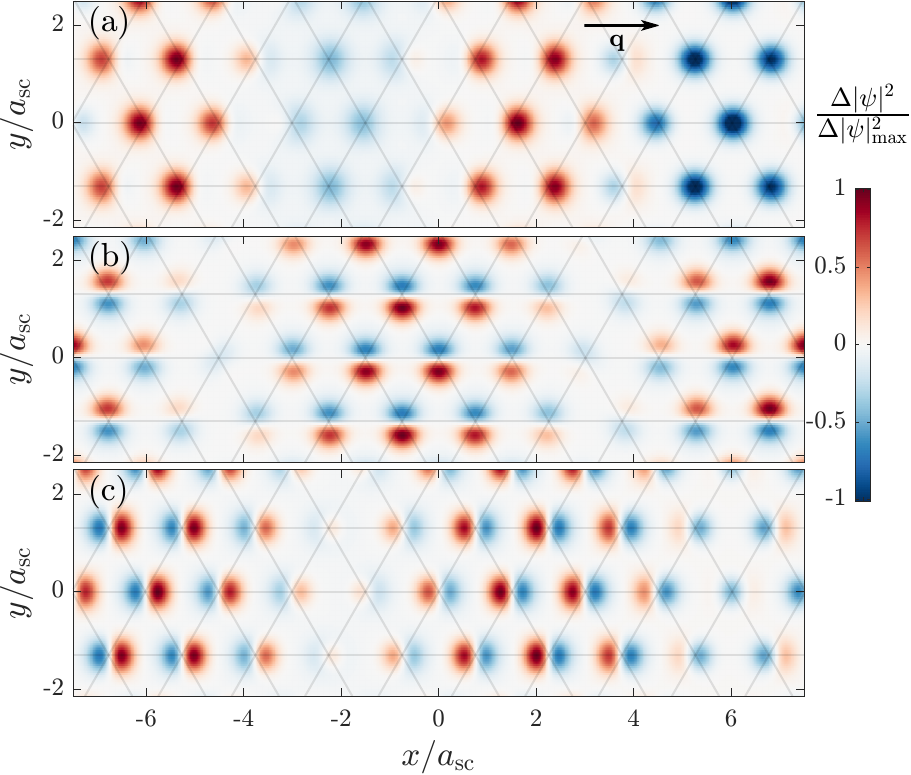}
	\caption{Effect of excitations from the lowest three gapless bands on the supersolid density profile. The change in density $\Delta|\psi|^2$ (divided by its maximum value) from the addition of an excitation in the (a) second sound, (b) transverse, and (c) first sound bands. The lines intersect at the locations of the density peaks of the ground state. Results for 2D soft-core mode using $\mathbf{q}=(0.4,0)/a_{\mathrm{sc}}$ and {$\Lambda=42.5$}. [cf.~Fig.~\ref{bandstructure_softcore}(c)].
	\label{dynamics_softcore}}
\end{figure}

In all results we see that the gapless excitation bands emerge from the $\Gamma$ point, and that these all have a linear dependence on the excitation wavevector near $\Gamma$. For the dipolar case shown in Figs.~\ref{bandstructure}(b) and (c), the lowest branch (close to $\Gamma$) is a transverse excitation of the crystal, which has no weight in the dynamic structure factor\footnote{Note that at low $q$ the first and second sound branches have small weight due to the density fluctuations being suppressed, but for the transverse excitation branch the weight vanishes at all $q$.}.  The next branch is the second sound or phase mode, which has a weak density contribution. These two lowest branches have similar speeds of sound (i.e.~slope near $\Gamma$), and as $\epsilon_{dd}$ increases the  second sound speed decreases and becomes slower than the transverse speed of sound [see subplot (d)]. The third gapless branch of the spectra is known as first sound and is a longitudinal density wave. This excitation branch rises much more steeply than the other two, i.e.~exhibiting a much higher sound speed. It is instructive to compare these results to the spectra along the three symmetry directions calculated for the 2D soft-core supersolid (also studied in Ref.~\cite{Macri2013a}). As shown in  Fig.~\ref{bandstructure_softcore}, the order of the three branches is preserved while varying $\Lambda$, with the gapless transverse branch always sandwiched in between the second sound mode and the first sound mode. In Fig.~\ref{dynamics_softcore}, we show three exemplar modes that have the same quasimomentum $\mathbf{q}$ but belong to the three different gapless energy bands. To highlight the character of each mode, we plot the change in density $\Delta|\psi|^2 = \mathcal{N}^{-1}|\psi + c_{\nu\mathbf{q}}u_{\nu\mathbf{q}}(\mathbf{x})-c_{\nu\mathbf{q}}^*v^*_{\nu\mathbf{q}}(\mathbf{x})|^2-|\psi|^2$, obtained by subtracting the ground state density from the perturbed density, normalized by $\mathcal{N}$ to the same value as the ground state. Here, we observe that the second sound branch (a) keeps the location of the density maxima fixed, but density changes along the direction of propagation by atom tunnelling between sites. The transverse sound branch (b)  causes a shearing of the lattice sites (i.e.~transversal motion). The first sound branch (c) exhibits longitudinal displacement of the lattice sites, consistent with a classical crystal excitation. Notice that, to help the visualization, the quasimomentum $\mathbf{q}$ of the selected modes is chosen along along the $x$-axis, which is equivalent to the $\Gamma-K$ symmetry direction in Fig.\,\ref{bandstructure_softcore}.
 
\section{Hydrodynamic description}\label{Sec:Hydro}

The features of the linear part of the excitation spectrum of a 2D supersolid, i.e., the low-energy region for small momenta ($q\ll\pi/a$), are well captured by a hydrodynamic description \cite{Son2005a,Yoo2010a}. This approach is based on long-wavelength perturbations of the ground state, obtained by applying small variations to the three fields associated with conserved quantities and broken symmetries: the change in average density $\delta\rho$, the displacement field $u_{i=\{x,y\}}$ that deforms the planar coordinates as $ {x_i}\to x_i^\prime=x_i+u_i$, and the superfluid phase field $\phi$.
Before writing the Lagrangian density for a 2D supersolid to obtain the three speeds of sound, it is useful to extract the elastic parameters of the system. Later, these will appear as coefficients of the Lagrangian. 

\subsection{Elastic parameters}\label{Sec:ElasticParams}
  The first parameter is the superfluid fraction determined by the energetic response of the ground state to changes in the superfluid velocity (see Ref.~\cite{Saslow1976a,Sepulveda2010a,Blakie2024a}). In general the superfluid fraction is a tensor given by
\begin{align}
f_{s,ij}=\frac{1}{m\rho}\frac{\partial^2\mathcal{E}}{\partial v_i\partial v_j},
\end{align}
where the indices $i,j=\{x,y\}$  denote the planar coordinates.
For the triangular ground state this tensor is isotropic, i.e. $f_{s,ij}=f_s\delta_{ij}$, and we can simply refer to the superfluid fraction as a scalar value $f_s$.  As a consequence, the average superfluid density $\rho_s = f_s\rho$, and the average normal density $\rho_n = (1-f_s)\rho$, are also isotropic quantities.

The other elastic parameters involve the dependence of the energy density on the areal density and lattice vectors. The second derivative of $\mathcal{E}$ with respect to the average density $\rho$ defines
\begin{align}
\alpha_{\rho\rho}= \frac{\partial ^2\mathcal{E}}{\partial \rho^2},
\label{alpharr}
\end{align}
which relates to the isothermal compressibility at constant strain:  
\begin{align}
\tilde\kappa=\frac{1}{\rho^2\alpha_{\rho\rho}}.\label{kappatilde}
\end{align}
The elastic tensor associated with the crystalline structure is given by
\begin{align}
C_{ijkl}=  \frac{\partial ^2\mathcal{E}}{\partial u_{ij}\partial u_{kl}} ,\label{alphauu}
\end{align}
 {where $u_{ij}=\frac{1}{2}(\partial_iu_j+\partial_ju_i)$ is the strain tensor arising from the displacement field $\mathbf{u}$.} 
We obtain this tensor by evaluating how the ground state energy density changes with the lattice vector deformations  $a_{\sigma={1,2},i}^\prime =(\delta_{ij}+u_{ij})a_{\sigma,j}$ (using the Einstein summation convention). We have verified that the elastic tensor for the triangular ground state is isotropic \cite{Landau1986}, being of the form 
\begin{align}
C_{ijkl}=\tilde\lambda\delta_{ij}\delta_{kl}+\tilde\mu(\delta_{ik}\delta_{jl}+\delta_{il}\delta_{jk}),\label{Cijkl}
\end{align}
where $\{\tilde\lambda,\tilde\mu\}$ are the Lam\'e parameters. We denote $\alpha_{uu}=\tilde\lambda+2\tilde\mu$, which is the diagonal element of the elastic tensor (i.e.~$C_{xxxx}=C_{yyyy}$), also known as the longitudinal or P-wave modulus. We also have a special interest in  the shear modulus $\tilde{\mu}$, which  is given by off-diagonal tensor elements, such as $C_{xyxy}$. \\
 Finally, we consider the density-strain coupling parameter given by the mixed partial derivative
\begin{equation}
    \alpha_{\rho u}=\frac{\partial^2 \mathcal{E}}{\partial u_{ii} \partial \rho}\,,
\end{equation}
which describes the coupling between changes in average density and the unit cell area.

\begin{figure}[htbp]
	\centering
	\includegraphics[width=3.4in]{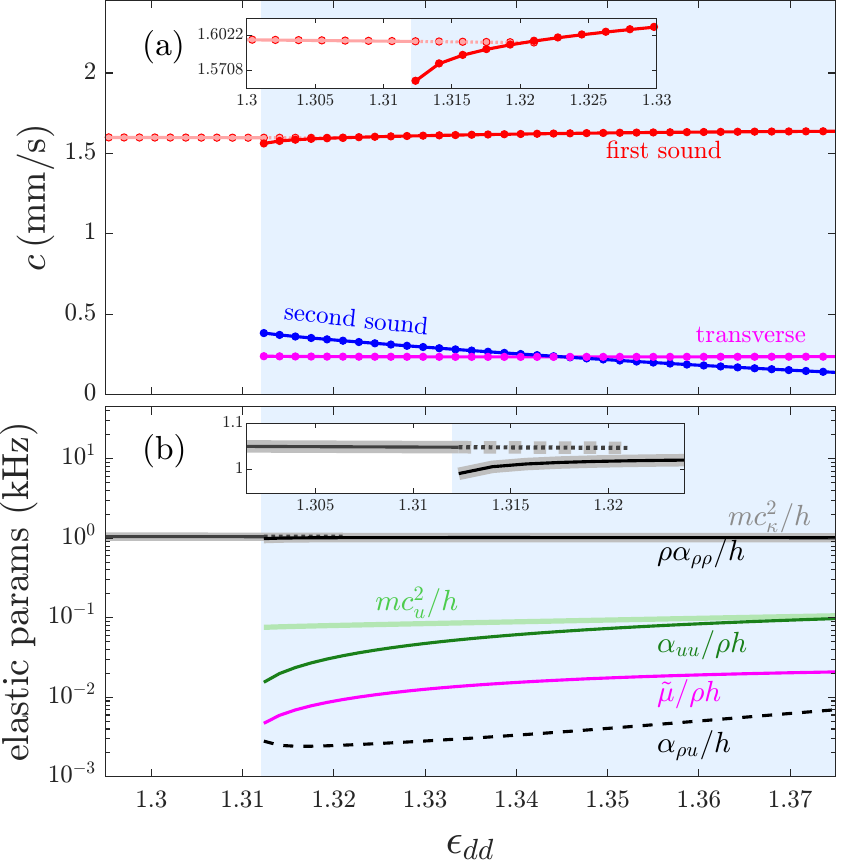}
    \caption{Elastic properties and speeds of sound for a planar dipolar BEC at the first-order transition to a supersolid.  (a) Speeds of sound. We show the speed of first sound in the uniform  (light red) and supersolid (red) phases,  and the speeds of transverse  and second sound in the supersolid phase (as labelled in subplot). Results from the BdG analysis are shown as lines, and the hydrodynamic predictions using the elastic parameters from the ground state calculations are shown as markers. (b) The elastic parameters characterising the dependence of the ground state energy on lattice and density changes. The characteristic energies $mc_u^2$ \eqref{cuu} (light green) and  $mc_\kappa^2$ \eqref{ckappa} (grey) are also shown for reference. In all subplots the uniform-to-supersolid transition point is indicated by the white to light-blue shaded area in the background. The uniform results are continued below the transition (where it is metastable) until it becomes dynamically unstable when the roton softens (at $a_s\approx99a_0$).	Insets to subplots (a) and (b) show the behaviour near the transition in detail.  The other parameters are the same as in Fig.~\ref{bandstructure}.
	\label{dipolarexcitations}}
\end{figure}

\begin{figure}[htbp]
	\centering
	\includegraphics[width=3.4in]{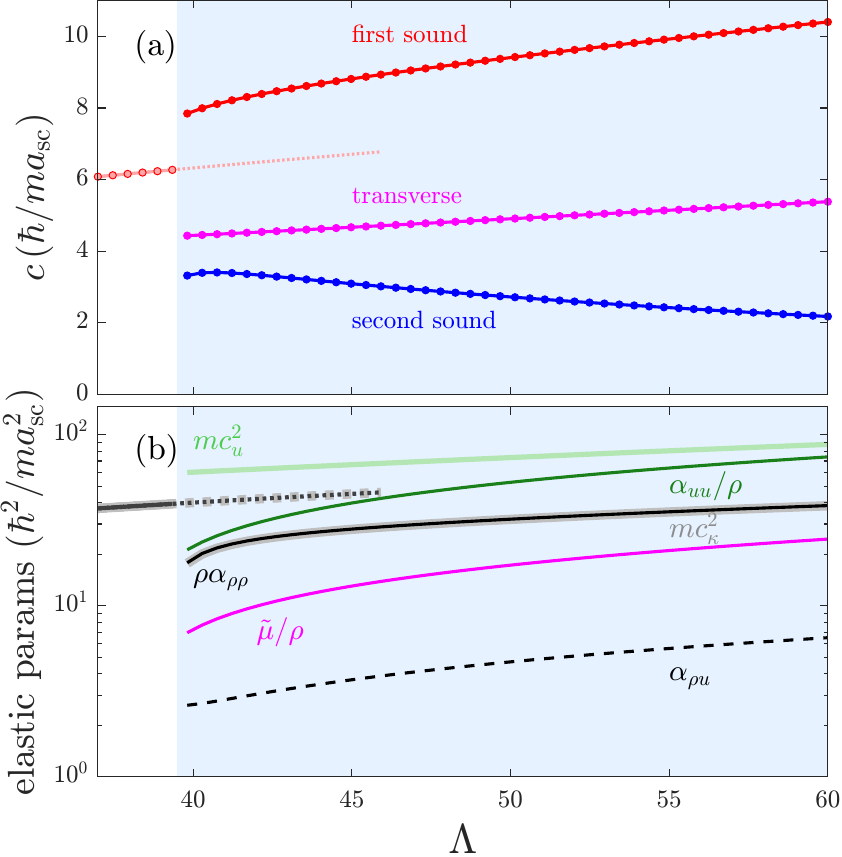} 
		\caption{Elastic properties and speeds of sound for a 2D soft-core BEC at the first-order transition to a supersolid. (a) speeds of sound and (b) elastic parameters. The color code and the quantities are the same as in Fig.\,\ref{dipolarexcitations}. The uniform results are continued below the transition (where it is metastable) until it becomes dynamically unstable when the roton softens (at $\Lambda = 46.30$).
	\label{softcoreexcitations}}
\end{figure}

\subsection{Hydrodynamic theory}
 {We now give the quadratic Lagrangian density for the fields $\delta\rho$, $u_i$, $\delta\phi$, and solve the Euler-Lagrange equations to extract the speeds of sound. The quadratic Lagrangian density for a supersolid, introduced by Yoo \textit{et al.}~\cite{Yoo2010a}, reads:
\begin{equation}
\begin{aligned}
\mathcal{L}_{\mathrm{SS}}^{\text {quad }}= & -\hbar\delta \rho \partial_t \phi
-\rho\frac{\hbar^2}{2m}\left(\partial_i \phi\right)^2-\alpha_{\rho u}\delta \rho \partial_iu_{i}\\
&-\frac{1}{2} \alpha_{\rho\rho}(\delta \rho)^2 +\frac{1}{2} m\rho_{n}\left(\partial_t u_i-\frac\hbar m\partial_i \phi\right)^2\\
&-\frac{1}{2} C_{ijkl} \partial_i u_{j} \partial_ku_{l}.\label{Lagrangian}
\end{aligned}
\end{equation} 
 From the solution of the  Euler-Lagrange equations   (see Appendix \ref{Sec:ELeq}), one gets the three speeds of sound:
\begin{align}
    mc_1^2= & \frac12(a_\Delta + \sqrt{\smash{a_\Delta^2} - 4b_\Delta}),\label{c1}\\
    mc_2^2= & \frac12(a_\Delta - \sqrt{\smash{a_\Delta^2} - 4b_\Delta}),\label{c2}\\
    mc_t^2= & \frac{\tilde\mu}{\rho_n},\label{ct} 
\end{align} 
where we have defined
\begin{align}
 a_\Delta = &\rho\alpha_{\rho\rho}-2\alpha_{\rho u} + \frac{\alpha_{uu}}{\rho_n},\\
b_\Delta = &\frac{\rho_{s}}{\rho_{n}}\left(\alpha_{\rho\rho}\alpha_{uu}-\alpha_{\rho u}^2\right)\,.
\end{align}
The quantities $c_1$, $c_2$ and $c_t$ represent the first, second and   transverse speeds of sound, respectively, and they are fully determined by the elastic coefficients and density of the system.}

\subsection{Hydrodynamic results}
In Figs.~\ref{dipolarexcitations} and \ref{softcoreexcitations} we consider properties of the dipolar and soft-core systems across the superfluid to supersolid phase transition. 
The speeds of sound extracted from linear fits to the $q\to0$ behavior of BdG energies calculated for the gapless branches are shown in Figs.~\ref{dipolarexcitations}(a) and \ref{softcoreexcitations}(a). The uniform superfluid state only has a single gapless branch: first sound  ($c_1$). In the transition to a supersolid there is almost no change in  $c_1$ for the dipolar case, while a significant upward jump in $c_1$ occurs for the soft-core case. We also note that in the dipolar system $c_1$ is higher relative to the other speeds of sound than in the soft-core system.  
The second sound speed $c_2$ is always much lower than first sound. Except close to the transition in the soft-core system, $c_2$  generally decreases as we go deeper into the crystalline phase. This reduction in $c_2$ is a sign of the reduced superfluidity. Indeed, sufficiently far into the crystalline phase we expect a transition to an isolated droplet crystal and the superfluidity will vanish, although this is beyond the theories we use here (see \cite{Buhler2023a}).

\begin{figure}[htbp]
	\centering
	\includegraphics[width=3.4in]{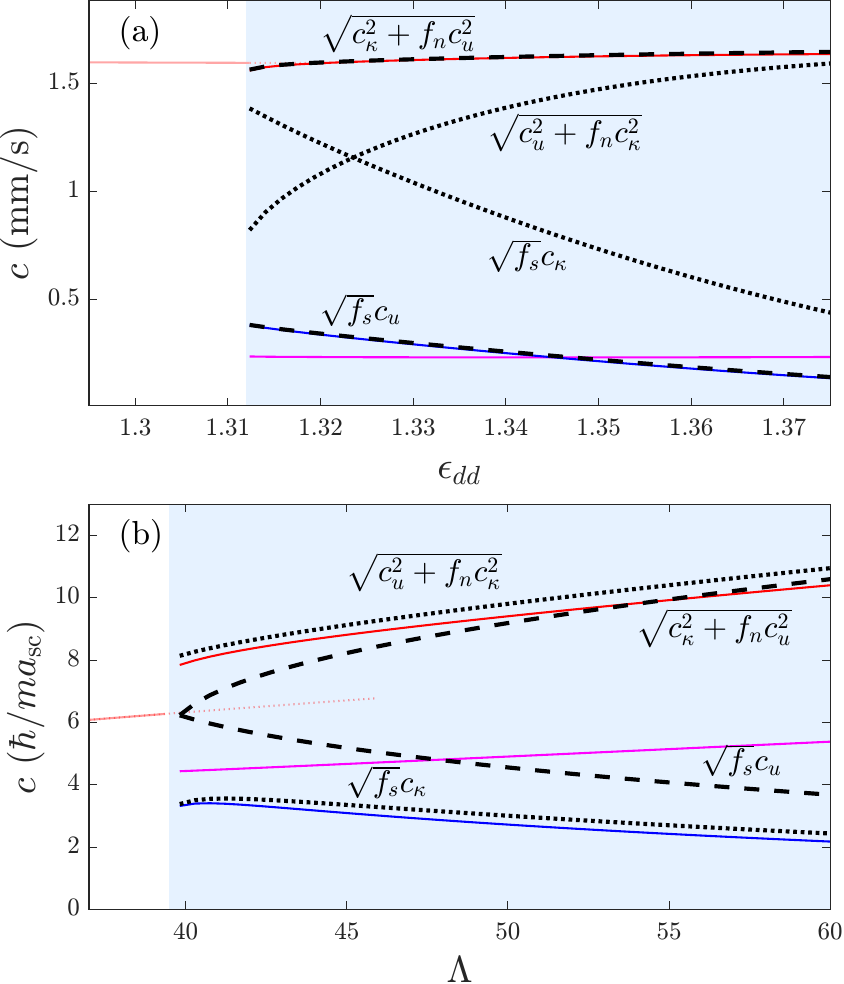}
	\caption{Comparison of first and second speeds of sound to limiting results for the (a) dipolar  and (b) soft-core results. Colored solid and dotted lines are the BdG results for the speeds of sound with the same colors as used in Figs.~\ref{dipolarexcitations}(a) and \ref{softcoreexcitations}(a). Black dotted and dashed lines indicate the rigid lattice and bulk incompressibility limiting results for $c_1$ and $c_2$.  
	\label{addanalysis}}
\end{figure}

The elastic parameters extracted from the ground state solutions are presented in Figs.~\ref{dipolarexcitations}(b) and \ref{softcoreexcitations}(b), with $f_s$ appearing in Figs.~\ref{bandstructure}(a) and \ref{bandstructure_softcore}(a). From these parameters we can evaluate the hydrodynamic predictions for the speeds of sound from Eqs.~(\ref{c1})-(\ref{ct}). These results are plotted as symbols in  Figs.~\ref{dipolarexcitations}(a) and \ref{softcoreexcitations}(a) and reveal excellent agreement with the BdG results.  
It is interesting to consider the role of the density-strain term $\alpha_{\rho u}$, as our results show this term is relatively small, and it is neglected in some treatments \cite{Hofmann2021a,Sindik2024a}. For the dipolar results, dropping this term shifts the hydrodynamic prediction for $c_2$ down by about $0.3\%$ and the $c_1$  prediction up by $0.4\%$. Including the density strain term gives results in better agreement with the BdG results, but both sets of results are almost indiscernible from the BdG results on the scales of our figures.
 
\subsection{Limiting behavior}
To gain a deeper understanding of the behavior of these systems it is useful to  introduce the characteristic speeds  $c_\kappa$ and $c_u$ \cite{Platt2024a}, defined as
 \begin{align}
 mc_{\kappa}^2 &\equiv \rho \frac{ \alpha_{\rho\rho}\alpha_{uu}-\alpha_{\rho u}^2}{\alpha_{uu}}=\frac{1}{\rho\kappa}, \label{ckappa}\\
 mc_{u}^2 &\equiv \frac {\alpha_{uu}}{\rho_n}\label{cuu}.
 \end{align}
 The first speed is associated with the system compressibility $\kappa$ \cite{Yoo2010a,Platt2024a} and the second is associated with the lattice elastic properties. In the uniform superfluid, with a single gapless excitation branch that dominates the low $q$ density response of the system, $c_{\kappa}$ describes the first sound\footnote{Here  $\alpha_{uu}$ and $\alpha_{\rho u}$ are both zero and Eq.~(\ref{c1}) reduces to $c_{\kappa}$.}. In the supersolid phase neither of these characteristic speeds correspond to any of the  speeds of sound, but they are useful for defining two regimes of behavior. For reference, we indicate these quantities compared to the other elastic parameters in Figs.~\ref{dipolarexcitations}(b) and \ref{softcoreexcitations}(b). We observe that $c_\kappa\gg c_u$ in the dipolar supersolid, whereas $c_\kappa\ll c_u$ in the soft-core case. This emphasizes the contrasting importance of compressibility and lattice elasticity in the properties of the two supersolids.

  \subsubsection{Bulk incompressible limit} 
 We take the bulk incompressible limit to be when $c_u\ll c_\kappa$, which is relevant to the dipolar supersolid. Within this regime $c_t$ is unaffected but  $c_1$ and $c_2$ become
 \begin{align}
 	c_1 & \to \sqrt{c_\kappa^2+f_nc_u^2},\\
 	c_2 & \to \sqrt{f_s}c_u,
 \end{align}
 where $f_n=1-f_s$ is the normal fraction. These two limits approximately describe the first and second sound velocity in the dipolar model, as shown in Fig.~\ref{addanalysis}(a).
 
 \subsubsection{Rigid lattice limit}
 The rigid lattice limit  occurs when $c_u\gg c_\kappa$. As noted before, this is the regime appropriate to the soft-core supersolid. Within this regime $c_t$ is unaffected but $c_1$ and $c_2$ become
 \begin{align}
c_1 & \to \sqrt{c_u^2+f_nc_\kappa^2},\\
c_2 & \to \sqrt{f_s}c_\kappa.
\end{align}
These two limits approximately describe the first and second sound velocity in the soft-core 2D model, as shown in Fig.~\ref{addanalysis}(b).

The extreme case $c_u\to \infty$ can be realised with a BEC loaded  into an optical lattice\footnote{This is an artificial supersolid, because translational invariance is not spontaneously broken.}. Here the lattice sites cannot move (cf.~Refs.~\cite{Chauveau2023a,Tao2023a}) and the system has a single gapless band corresponding to second sound.

\section{Outlook and Conclusions}\label{Sec:Conclusions}
 In this paper we have contrasted the excitation spectra of $D=2$ supersolids arising in two different systems.   
 We have focused on the three gapless excitation branches and their associated speeds of sound.  Our work demonstrates that hydrodynamic theory provides an accurate prediction of the speeds of sound based only on a set of generalized elastic parameters that are determined from ground state calculations. We find that the dipolar system is dominated by its compressibility ($c_\kappa\gg c_u$) whereas the soft-core system is in the rigid lattice regime ($c_\kappa\ll c_u$). As a result the first sound speed of the dipolar system is significantly larger than the other speeds of sound, and there is barely any jump in the first sound speed at the superfluid to supersolid transition. A recent proposal has suggested a method for  measuring the  long-wavelength excitation frequencies, and longitudinal speeds of sound in a 1D supersolid using a periodic optical potential \cite{Sindik2024a}. This method can be directly extended to the 2D supersolid.

 Our results will provide a basis for better understanding the equilibrium and dynamical properties of higher dimensional supersolids, which are being explored in experiments with dipolar gases. We have also calculated the excitations and compared to the hydrodynamic theory for a higher density dipolar BEC ($\rho=0.08/a_{dd}^2$) and  find results that are qualitatively similar to those presented here, except that $c_\kappa$ is slightly larger.  An interesting remaining question is to extend our analysis to even higher densities, where the dipolar supersolid transitions to a stripe state or a hexagonal state. An interesting aspect of the stripe state that its superfluid fraction and elastic tensor will become anisotropic. Such high densities will be difficult to reach in experiments with magnetic gases, but these parts of the phase diagram might be accessible to polar molecules gases which have strong dipole-dipole interactions \cite{Schmidt2022a,Bigagli2024a}.  This presents a rich playground for studying superfluidity in the presence of crystalline order.

\emph{Note added.} Recently the preprint Ref.~\cite{Rakic2024a} appeared, which also considers the elastic properties of soft-core supersolids.

\section*{Acknowledgments}
\noindent  DB and PBB acknowledge use of New Zealand eScience Infrastructure (NeSI) high performance computing facilities and support from the Marsden Fund of the Royal Society of New Zealand. We acknowledge useful conversations and comparisons with  P.~Senarath Yapa  and T.~Bland. E.P.~acknowledges support by the Austrian Science Fund (FWF) within the DK-ALM (No.\,W1259-N27).

\appendix
\section{Planar dipolar BEC}\label{Sec:ModelDBEC} 
Here we discuss the details of our model of a planar dipolar BEC of magnetic atoms.
The atoms are free  move in the  $xy$-plane and the planar kinetic energy is given by
\begin{align}
T_\mathbf{v}=\frac{(-i\hbar\bm{\nabla}_\perp+m\mathbf{v})^2}{2m},
\end{align}
where $\bm{\nabla}_\perp$ is the 2D gradient and we have allowed for a superfluid velocity $\mathbf{v}$ (cf.~\cite{Roccuzzo2019a,Aleksandrova2024a}). In the $z$-direction the single particle Hamiltonian includes harmonic confinement
\begin{equation}
	H_z= -\frac{\hbar^2}{2m}\frac{\partial^2}{\partial z^2} + \frac12 m\omega_z^2z^2,
\end{equation}
with $\omega_z$ being the angular frequency.
The magnetic dipole moments of the atoms are taken to be polarized along $z$ by a bias field and the interactions are described by the potential 
\begin{equation}
	U(\mathbf{r}) = \frac{4\pi a_s\hbar^2}{m}\delta(\br) + \frac{3\add\hbar^2}{m r^3}\left(1-3\frac{z^2}{r^2}\right),
\end{equation}
where $ \textbf{r}= \textbf{x}- \textbf{x}'$ is the relative separation between the particles. Here $a_s$ is the$s$-wave scattering length,  $\add = m\mu_0\mu_m^2/12\pi\hbar^2$ is the dipole length, and $\mu_m$  is the atomic magnetic moment.
The ratio $\edd=\add/a_s$ characterises the relative strength of the dipole-dipole to $s$-wave interactions and  when this parameter is sufficiently large the ground state undergoes a transition to a crystalline state with spatial modulation in the $xy$-plane.

The eGPE energy functional for this system is  
\begin{align}
E &= \int_{\mathrm{uc}} d\bx\, \psi^*\left(T_\mathbf{v}+H_z+\tfrac12\Phi   +\tfrac25\gammaQF|\psi|^3\right)\psi,\label{Efunc}
\end{align}
where we have introduced the effective potential
 \begin{align}
\Phi(\bx)=\int d\bx'\,U(\bx-\bx')|\psi(\bx')|^2,
\end{align}
and the effects of quantum fluctuations are described by the term with coefficient $\gammaQF = \frac{128\pi\hbar^2}{3m}a_s\sqrt{\frac{a_s^3}{\pi}}\mathcal{Q}_5(\edd)$, where $\mathcal{Q}_5(x)=\Re\{\int_0^1 du[1+x(3u^2 - 1)]^{5/2}\}$ \cite{Lima2011a}.
Because the system is infinite in the $xy$-plane we restrict the spatial extent of the wavefunction to the unit cell (uc) defined by the direct lattice vectors $\{\mathbf{a}_1,\mathbf{a}_2\}$ and impose periodic boundary conditions. To accurately calculate $\Phi(\bx)$ in the unit cell we employ the $z$-cutoff truncated interaction potential introduced in Ref.~\cite{Ronen2006a}.

The average density condition enforces the following normalization constrain on the wavefunction 
\begin{align}
\int_{\mathrm{uc}}d\mathbf{x}\,|\psi(\mathbf{x})|^2=\rho A,
\end{align}
where  $A=|\mathbf{a}_1\times\mathbf{a}_2|$ is the area of the unit cell.
The ground state is determined for a specified value of $\rho$ and $\mathbf{v}=\mathbf{0}$ by minimising the energy density 
\begin{align}
\mathcal{E}(\rho,\mathbf{a}_1,\mathbf{a}_2,\mathbf{v})\equiv E/A.
\end{align} 
with respect to the unit cell parameters $\{\mathbf{a}_1,\mathbf{a}_2\}$. For the purpose of computing the elastic parameters, we are also interested in small changes in the parameters $\{\rho,\mathbf{a}_1,\mathbf{a}_2,\mathbf{v}\}$ (i.e.~to make finite difference derivatives) from the ground state values. In this case the energy density is minimised with respect to $\psi$, but with the other parameters specified.

\section{2D soft-core condensate}\label{Sec:ModelSCBEC}
The 2D soft-core model is for a BEC of atoms in the $xy$-plane and interacting with a potential of the form
$U_\mathrm{sc}(\mathbf{r})=U_0\theta(a_\mathrm{sc}-|\mathbf{r}|)$, where $a_\mathrm{sc}$ is the soft-core radius, $U_0$ is the potential strength, and $\theta$ is the Heaviside step function. The meanfield energy functional for this system is similar to the dipolar case, but with out any $z$-direction or quantum fluctuations
\begin{align}
E &= \int_{\mathrm{uc}} d\bx\, \psi^*\left(T_\mathbf{v}+\tfrac12\Phi_{\mathrm{sc}}\right)\psi,\label{Efuncsc}
\end{align}
where
 \begin{align}
\Phi_{\mathrm{sc}}(\bx)=\int d\bx'\,U_{\mathrm{sc}}(\bx-\bx')|\psi(\bx')|^2.
\end{align}
In this system it is convenient to adopt $a_\mathrm{sc}$ as the unit of length and $\hbar\omega_0=\hbar^2/ma_{\mathrm{sc}}^2$ as the unit of energy, and to define the dimensionless interaction  parameter
\begin{align}
\Lambda = \frac{m\pi \rho a_{\mathrm{sc}}^4U_0}{\hbar^2} .
\end{align} 
The normalization condition, definition of energy density, and procedure to obtain ground states is then the same as described for the dipolar case.

It is worth noting that as discussed by Macr\`i {\it et al.}~\cite{Macri2013a}, for any finite value of $\Lambda$ the meanfield description of 2D soft-core system will be valid for sufficiently high density $\rho$. There have also been comparisons validating the meanfield 2D soft-core model against quantum Monte Carlo results in Ref.~\cite{Macri2013a} (also see \cite{Cinti2014a}).

\section{Bogoliubov-de Gennes equations}\label{Sec:BdG}

The Bogoliubov-de Gennes (BdG) framework describes the quasiparticle excitations  about a stationary solution\footnote{Here we consider solutions for $\mathbf{v}=0$ where  $\psi$ can be taken to be real.}  $\psi(\mathbf{x})$. This solution satisfies the time-independent GPE $\mathcal{L}\psi=\mu\psi$, where
\begin{align}
\mathcal{L}&= T_\mathbf{0} + \Phi_{\mathrm{sc}},\quad&\mbox{(soft-core)}\\
\mathcal{L}&= T_\mathbf{0}+H_z + \Phi   +\gammaQF\psi^3, \quad&\mbox{(dipolar)}
\end{align}
and $\mu$ is the chemical potential.

The quasiparticle modes can be determined by linearising the time-dependent evolution, $i\hbar\dot{\Psi}=\mathcal{L}\Psi$, with an expansion of the form given in Eq.~(\ref{psipert}).
This leads to the BdG equations
\begin{align}
\begin{pmatrix} 
      \mathcal{L}+X-\mu & -X \\
      X & -(\mathcal{L}+X-\mu) \\
   \end{pmatrix} \begin{pmatrix} u_{\nu\mathbf{q}} \\ v_{\nu\mathbf{q}}  \end{pmatrix} =\hbar\omega_{\nu\mathbf{q}}\begin{pmatrix} u_{\nu\mathbf{q}} \\ v_{\nu\mathbf{q}}  \end{pmatrix},
\end{align}
where $u_{\nu\bq}$ and $v_{\nu\bq}$ have the Bloch form $f_{\nu\bq}(\bx)=\bar{f}_{\nu\bq}(\bx)e^{i\bq\cdot\bx}$ with $\bar{f}_{\nu\bq}(\bx)$ periodic in the unit cell, and $X$ is defined so that 
\begin{align}
\!Xf&=\!\psi \!\int\!d\mathbf{x}^\prime U_{\mathrm{sc}}(\mathbf{x}\!-\!\mathbf{x}^\prime)f(\mathbf{x}^\prime)\psi(\mathbf{x}^\prime),\quad&\mbox{(soft-core)}\\
\!Xf&=\!\psi \!\int\!d\mathbf{x}^\prime U(\mathbf{x}\!-\!\mathbf{x}^\prime)f(\mathbf{x}^\prime)\psi(\mathbf{x}^\prime) \ldots \nonumber\\
&+\frac32\gamma_{\mathrm{QF}}\psi^3f.\quad&\mbox{(dipolar)}
\end{align}

\section{Euler Lagrange equations}\label{Sec:ELeq}

From Eq.~(\ref{Lagrangian}) we obtain the Euler-Lagrange equations to describe the evolution of the hydrodynamic fields
\begin{align}
	\hbar\partial_t\phi &= -\alpha_{\rho\rho}\delta\rho-\alpha_{\rho u} \partial_i u_i,\\
	m(\partial_t\delta\rho-\rho_n\partial_{ti}u_i)  &= -\hbar\rho_s\partial^2_i\phi,\\
	\rho_n(m\partial_t^2 u_i -\hbar\partial_{ti}\phi)  &=  \alpha_{\rho u}\partial_i\delta \rho + C_{ijkl}\partial_{jk}u_l.\label{EL3}
\end{align} 
We can further decompose $\mathbf{u}=\bu_t+\bu_l$ in to transverse ($\bu_t$ with $\nabla\cdot\bu_t=0$) and longitudinal ($\mathbf{u}_l$ with $\nabla\times \bu_l = 0$) parts. Using this decomposition and Eq.~(\ref{Cijkl}), the last Euler-Lagrange equation can be written as 
\begin{align}
    \rho_n(m\partial^2_t\bu_l-\hbar\partial_t\nabla\phi) &= \alpha_{\rho u}\nabla\delta\rho + \alpha_{uu}\nabla^2\bu_l,\\
    \rho_nm\partial^2_t\bu_t &=  \tilde\mu\nabla^2\bu_t.
\end{align}
The normal mode solutions of these equations are of the form $X=X_we^{i(\mathbf{q}\cdot\mathbf{x}-\omega t)}$, where $X=\{\delta \rho,\phi,u_l,u_t\}$  and $X_w$ is the excitation amplitude (also see \cite{Yoo2010a,Platt2024a}). Three non-trivial solutions can be found with dispersion relations of the form $\omega=c q$, yielding the speeds of sound given in Eqs.~(\ref{c1})-(\ref{ct}).

\end{document}